# Driving an NP-Complete problem with Combinatorial Decomposition to generate a unique and irreversible bitstring from a single integer seed value.

S.Christmas, R.Leidich

Jan 7[th], 2013

## Abstract

Generation of an (arbitrarily) long string of bits unique to a given finite-length numerical "seed" is of great value in the field of random number generation, computer simulations, and other areas of computer science. Extending this idea such that the bitstring cannot be reverse-engineered to recover the original seed value extends the value of such a system to the field of cryptography, as the string can be used directly as an encryption mask, or as the input to some other cryptographic function. The longer the string that can be generated, the closer the system would come to the ideal cryptographic case of the "One Time Pad"[1].

In this paper we propose a scheme for taking an initial seed (nominally a 128-bit integer, but not restricted to such), and expanding this into a unique bitstring of a length determined by a limit cycle that makes it useful in practical applications. We utilize novel mathematical concepts such as combinatorial decomposition to turn the seed value into unique rotations of a pre-defined table, which are operated on via destructive functions such as exclusive-OR (XOR) and add-with-carry (ADC) to eventually produce the unique bitstring.

We assert that the process of iterating the XOR or ADC operation conforms to the known NP-complete problem known as Subset-Sum, meaning that the reversal of the process that produced the bitstring is tantamount to solving the NP-Complete Subset-Sum problem, which in turn is less efficient than the brute-force method of testing every seed value until the corresponding bitstring is found, making the system highly relevant in a cryptographic context.

### 1. The Circular Array (JP-Table)

The heart of the system is a circular array of randomly distributed 0's and 1's, hereafter referred to as the Janssen-Palmer (JP) Table, after the individuals who inspired the research. This is a string of bits, containing randomly distributed 0's and 1's in equal numbers. The string is considered to be a "ring", so that the first and last bits are effectively neighbours. We do not define a standard size for the JP-Table, but we suggest $2^{16}$ = 65536 bits offers the best tradeoff between speed and average cycle length of the bitmask when tested on standard desktop computer hardware.

A graphical representation of the JP-Table is shown here. Note also a feature of the table is the predetermined "Start Point", which is critical to the system.

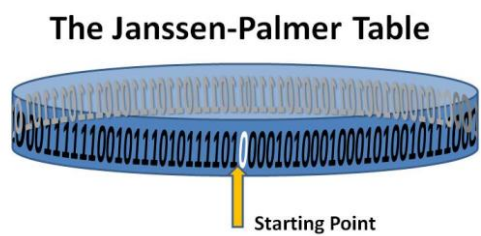

[Figure 1: The JP-Table]

If the system is to be used as a method of pseudo-random number generation (PRNG) or basis for a code path in computer simulation, the JP-table should be re-generated as often as deemed necessary, using another PRNG with a long limit cycle such as the Mersenne Twister[2]. It may seem counter-intuitive to seed one PRNG with another, but it will become clear that any bias of the PRNG used to seed the JP-Table is quickly whitened out after a few iterations of the process, except in the most extreme cases.

If the method is to be used as a basis for a cryptographic system, the JP-Table should be standardized to allow for common implementations. However, the secrecy of such a system would not depend on the secrecy of the JP-Table, which would be make public for the aforementioned reason of allowing common implementations.

## 2. Tumblers

A key part of the system involves rotations of the JP-Table as follows. Take the sample circular array (the JP-Table) shown in Figure 1, and imagine rotating it from the highlighted starting point by 27 bit positions (in any direction, but the direction of rotation should be the same from there on out).

This results in effectively a new table, equivalent to the original table rotated with respect to the starting bit position. The decimal number 27 is thus known as a "Tumbler Value" corresponding to this new rotated table, and the actual rotated table is the "Tumbler", as it serves a purpose analagous to the tumblers of the locks on a briefcase. Thus, given the JP-Table size of $2^{16}$ bits, we have $2^{16}$ possible rotation positions, if we include the original table as a Tumbler with Tumbler Value 0. This we have 65536 16-bit Tumblers and Tumbler Values, numbered from 0 to 65535.

The system will generate a number of such Tumblers Values, and thus a number of rotated JP-Tables, which one can image as being stacked on top of one another, with all rotations and bit positions relative to the "start point". Then, an exclusive-OR (or alternately an add-with-carry) operation is performed "vertically" on all the bits (clearly in pairs of stacked tables) to form a new single "Ring Table", which will not equal the original JP-Table, but will necessarily be of the same size. Figure 2 shows this in more detail below.

To explain further, say we are given input integers from some source of 341, 573, and 1,204, which we will interpret as Tumbler Values. The operation involves first rotating the original JP-Table 341 places to the right (right being an arbitrary choice that we will now take as convention) and storing it. Then, the *original* JP-Table (*not* the new rotated table formed by the 341-bit rotation) is rotated 573 places to the right and also stored. Finally, the original table (again, not the previously rotated table) is rotated 1204 places, and all three rotations are XOR'd together (in pairs, so row 1 and 2, then the result of that with row 3) to form a new, single, circular array, the same size at the original JP-Table. Furthermore, we imagine all of these intermediate tables maintaining their alignment in space, so there remains a universal "start point" that aligns with the start point from which all rotations occurred, and this start point is preserved in the new circular array also.

Given that in the standard implementation the number of Tumblers used to form the new circular array is not known (unlike in this example where we know we have three Tumblers), if one is given only the *new* circular array and the original JP-Table, knowing exactly which Tumblers were used to form the new array is an adaptation of the Subset Sum problem, which is known to be NP-Complete[3]. This is true in a loose sense when the Tumblers are XOR'd together, and explicitly so when the add-with-carry (ADC) operation is used. The advantage to using XOR is that this part of the process can be made fully parallel assuming the hardware on which it is run allows for such an implementation. Using ADC sacrifices some of the parallel nature of this part of the scheme, but offers a more mathematical strictness to Subset Sum.

## 3. The Generic Iteration Process

Given that we have performed the process of rotating the JP-Table N times according to the (number of) and (value of) a set of integers, and performing a vertical XOR or ADC operations to produce a new circular array of 65536 bits, we can begin at the preserved start point and count off 16-bit chunks from the newly created circular array. This gives a whole new set of 16-bit integers of values between 0 and 65535 that we can interpret as new Tumbler Values (as well as other data structures that we will return to later). This whole operation can then be repeated, with the new Tumblers describing further rotational components of the *newly* created circular array.

We note here (and explain further later) that we cannot allow repeated Tumbler Values in any given iteration. This is referred to as "Repeated Tumbler Filtering" and does not need to occur every time this process is iterated, and so is included as a separate step in the process. However, when it is done, it is merged into the process described in this section, and is very simple. When a Tumbler Value is read off from the new circular array, it is compared with previous values. If there is a match, that new Tumbler Value is ignored, and the system moves to the next 16-bit chunk and reads a new value. Given the number of Tumbler values is M and the Table size is L, we always have $M <<< L$ and thus the chances of not being able to find M unique Tumbler Values in L is vanishingly small, and not a cause for concern.

Typically, we will read off the same number of Tumbler Value as the number of Tumblers we began with, and this will be covered in the next section, but it should be noted here that MORE than the initial number of Tumbler Values can be read from this new table, if Tumbler Count Expansion is needed for additional cryptographic security (again, this is described further later in the paper).

Note that after the initial "round" of rotations and XOR or ADC, the initial JP-Table is no longer used, as it is the newly formed circular array that is propagated through the system, being re-formed on every iteration. This iterative process is described here pictorially:

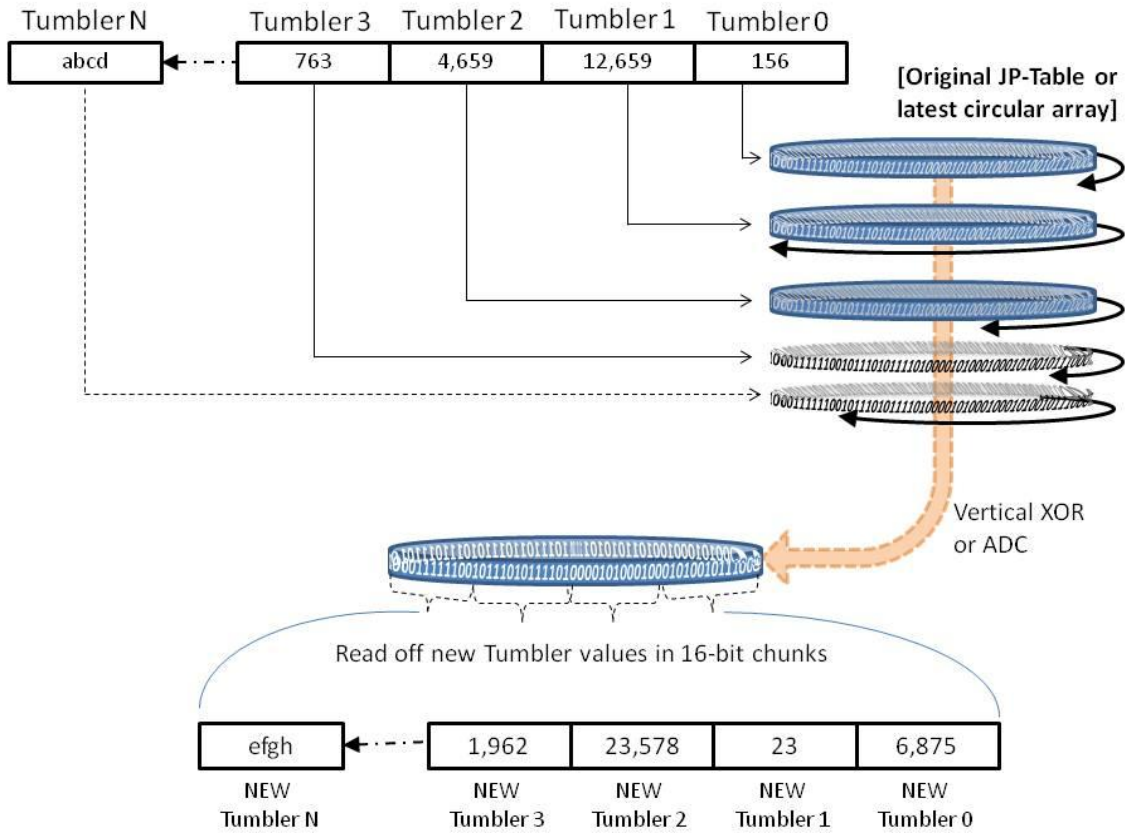

[Figure 2 : Iteration]

This iterative process is hereafter referred to as "Churning". Note that this procedure is always performed on the most recent incarnation of the circular array and not the original JP-Table, aside from the first iteration in which only the original JP-Table is available. Recall also that each rotation implied by the Tumbler Values in any given iteration is a rotation of the latest circular array from the zero-bit-position start point, not the rotation implied by the previous Tumbler Value. In other words, each iteration round starts with a "Master Ring" which is the original JP-Table in the first iteration, and the circular array produced by the previous iteration in all other cases. The above "Churning" process will hereafter be represented by the following symbol:

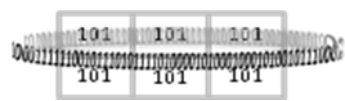

[Figure 3 : The Churning symbol]

## 4. Forming the initial Tumbler set from the initial seed: Combinatorial Decomposition.

The essence of Combinatorial Decomposition, that we introduce in this paper, is that any integer can be represented as a sum of combinations of a decrementing counter variable ranging from M down to 1, decrementing by 1 for each new combination, i.e. We can write the value of any integer P such that:

$$P = (m_0 \text{ nCr } M) + (m_1 \text{ nCr } M-1) + (m_2 \text{ nCr } M-2) + \ldots (m_{M-1} \text{ nCr } 1)$$

in concert with the conventions that $(n \text{ nCr } (n+1)) = 0$ and $(n \text{ nCr } n) = 1$, in which M is the counter variable and the $m_n$'s are integers that correspond uniquely to the value of the initial integer (the initial seed in this scheme), P.

The objective of combinatorial decomposition is that the initial seed is turned into a set of integers (the $m_n$'s above) which will then be interpreted as initial Tumblers. We add the condition that no two Tumblers are repeated in any given iteration (i.e during any given churn process). The reason for not allowing repeated Tumblers is that (n) XOR (n) = 0, which would weaken the computational complexity of reversing the implied bitstring back to the original Tumbler set (and then the initial seed value), which would be disastrous when used in a cryptographic context. Note that a repeated Tumbler does not explicitly allow this reverse engineering, but merely reduces the computational complexity of doing so by a factor of effectively two Tumblers.

Before going into the algorithm for combinatorial decomposition, some examples will be demonstrated with small numbers to aid in understanding the process.

Consider a very small hypothetical JP-Table, with only 6 bits: [0 1 1 0 1 0]. Remember this is really a ring, so the two end 0's are actually next to each other. Here are the possible rotations of the table (which we standardize as rotating *towards* the least-significant bit), each of which described by a unique Tumbler, 1-5. We will label them [$m_0$] for the first rotation (of zero places, i.e. the original table, so $m_0 = 0$) through to [$m_5$], which is a full rotation aside from one place (another rotation would bring it back to the beginning, giving a modulo-5 environment), and is equivalent to a rotation of one place in the direction *away* from the least-significant bit, or more simply $m_5 = 5$. Although it is clear that $m_x = X$, we keep the $m_n$ convention to maintain generality in the discussion.

$$\begin{array}{cccccc} 011010 & 110100 & 101001 & 010011 & 100110 & 001101 \\ m_0 & m_1 & m_2 & m_3 & m_4 & m_5 \end{array}$$

In order to generate a unique initial Tumbler set from the initial seed, we must identify a minimal subset of these Tumblers, which preserves all of the information in the initial seed. So, for instance, this could be [$m_0,m_3,m_5$] or [$m_2,m_3,m_4,m_5$] or just [$m_4$]. First, we need to determine how many Tumblers we will need, at minimum. Specifically, we need to find the *minimum* number of Tumblers, M, such that

$$[\textit{number of possible Tumblers (=JP-Table size)}] \text{ nCr } [M] > P$$

Where (nCr) is the standard combination function and P is the initial seed expressed as an integer. The following example should help to explain why this is the case, and it might help to re-state that we are seeking to expand 'P' according to

$$P = (m_0 \text{ nCr } M) + (m_1 \text{ nCr } M-1) + (m_2 \text{ nCr } M-2) + \ldots (m_{M-1} \text{ nCr } 1)$$

Given the preceding example of a hypothetical 6-bit JP-Table, and a (randomly chosen) initial seed of P = 18, we can show that the minimum number of Tumblers required (in this small example) is M=3, since

6 nCr **2** = 15 (less than P, therefore not acceptable) and

6 nCr **3** = 20 (greater than P as required).

Thus M = 3, i.e. we need at least 3 Tumblers to represent the full information implied by the initial seed.

So we know that we need to select 3 Tumblers from [$m_0…m_5$] above in such a way that there is a unique Tumbler set corresponding to every P and no Tumbler is repeated, since (a) XOR (a) = 0 as mentioned above. We also desire an algorithm that finds the Tumblers in descending order. This helps to ensure that no Tumblers are repeated without the use of $O(N^2)$ lookback tables and the like.

When used in any practical application, be it for a PRNG or in a cryptographic context, we desire at minimum more than one Tumbler, and work on the assumption that more Tumblers provides better results (we detail the minimum recommended number of Tumblers in various applications at the end of the paper). The above formula provides only the *minimum* number of Tumblers required to represent an integer with Combinatorial Decomposition from a purely mathematical point of view. We will typically require that we represent P with more than the mathematical minimum number of Tumblers.

How is this achieved? We will show how to handle this situation by example. Assume an initial seed value of P = 56 and we have somehow determined that to conform to an assumed level of cryptographic security in the resulting bitstring that we require 5 Tumblers.

The above formula shows that we only need at minimum one Tumbler, and that would be $m_n = 8$, since

8 nCr 5 = 56

And we have our value of P already. Since we have defined that we need five Tumblers to iterate through the process to provide the desired level of complexity, we use the conventions that (n) nCr (n+1) = 0 and (n) nCr (n) = 1 to give the following result:

P = (**8** nCr 5) + (**3** nCr 4) + (**2** nCr 3) + (**1** nCr 2) + (**0** nCr 1)

P =     56    +    0   +    0   +    0   +    0

P = 56

Which gives the Tumbler set (in bold underlined numerals above) of [8 , 3, 2, 1, 0]. Hence we have been able to use the prescribed five Tumblers to again give our seed value of P = 56, and we interpret these five Tumbler values as:

[*8* , 3, 2, 1, 0]

So we now return to our example of a 6-bit JP-Table, JP = [0 1 1 0 1 0], and the initial seed of P = 18. We have already shown that in this case M = 3, thus we need at least three Tumblers. It is helpful at this stage to again restate the equation

P = ($m_0$ nCr M) + ($m_1$ nCr M-1) + ($m_2$ nCr M-2) + …… ($m_{M-1}$ nCr 1)

which in this case simplifies to

P = ($m_0$ nCr 3) + ($m_1$ nCr 2) + ($m_2$ nCr 1)

Since M = 3. We need to then find the largest possible $m_0$ such that ($m_0$ nCr 3) is not greater than P (else the above equation is immediately violated and we clearly aren't finding P). This can be found in a number of ways, including a brute force search in which $m_0$ is increased from 2 (because we know that we need space for at least 2 lesser Tumblers) until ($m_0$ nCr 3) > P, then subtracting from P, and repeating an analogous process for $m_1$ and $m_2$, ultimately leaving 0 on the left side of the equation. Our proposed scheme employs a more efficient method, which is a binary search for the descending $m_n$'s over their allowed values, converging first on $m_0$, then $m_1$, etc.

For the brute-force case, assuming we find this $m_0$ such that ($m_0$ nCr 3) = X, we thus have:

$$P - X = (m_1 \text{ nCr } 2) + (m_2 \text{ nCr } 1)$$

The object is to find the largest X, and thus the largest $m_0$, such that (P-X) is whole. This process can be iterated until the entire initial seed is found. Using our example, we find the Tumblers as:

$$[\mathbf{\textit{5, 4, 2}}]$$

And this can be verified as follows:

$$(\mathbf{\textit{5}} \text{ nCr } 3) + (\mathbf{\textit{4}} \text{ nCr } 2) + (\mathbf{\textit{2}} \text{ nCr } 1) = 18$$
$$10 \quad + \quad 6 \quad + \quad 2 \quad = 18$$

We can also show that in the case of forcing the use of ten Tumblers rather than the mathematical minimum of three, we find the combinations as:

(11 nCr 10) + (9 nCr 9) + (8 nCr 8) + (7 nCr 7) + (6 nCr 6) + (5 nCr 5) + (4 nCr 4) + (3 nCr 3) + (1 nCr 2) + (0 nCr 1)

$$11 + 1 + 1 + 1 + 1 + 1 + 1 + 1 + 0 + 0 = 18$$

Which implies the M = 10 Tumbler set of:

$$[11, 9, 8, 7, 6, 5, 4, 3, 1, 0]$$

Thus, we have shown how to use the system of combinatorial decomposition to turn the initial seed into a unique and nonrepeating set of Tumbler Values, known as the "initial Tumbler set".

5. **Binary Search for Tumbler Sets**

We show here the more efficient binary search for Tumbler Sets.

In the case of the JP-Table, we know that $m_0$ must be on [M-1, $2^{16}$-1]. Any less than M-1 and there can't possibly be M Tumblers, or greater than $2^{16}$-1 and it doesn't represent a unique rotation of the JP-Table. So we execute a binary search over this domain to find the greatest ($m_0$ nCr M) that is not greater than P. For $m_1$, we can further reduce the domain to [M-2, $m_0$-1]. The lower bound is lower as we have fewer Tumblers remaining to find (since we subtracted X=($m_0$ nCr M) from P, leaving the left side as P-X) and the upper bound is one less than the $m_0$ just found as we are looking for the *next lesser* Tumbler. The process is then iterated, thus this approach to combinatorial decomposition improves upon brute force.

### 6. A Real-World Example

The above examples use small numbers to illustrate the theory involved, but are clearly not examples of a realistic or useful system by any means. In a cryptographic context, initial seeds would nominally be cryptographic keys of the order of $2^7$ to $2^9$ bits, along with the recommended JP-Table size of $2^{16}$ bits. Thus a typical seed value and corresponding Tumbler set would look as follows:

Initial seed in Hex format: **FEDCBA9876543210FEDCBA9876543210**

Combinations: (**32286** nCr 10)+(**32188** nCr 9)+(**31273** nCr 8)+(**24609** nCr 7)+(**24444** nCr 6)+

(**22362** nCr 5)+(**21029** nCr 4)+(**18123** nCr 3)+(**11302** nCr 2)+(**7367** nCr 1)

Tumblers: [**32286, 32188, 31273, 24609, 24444, 22362, 21029, 18123, 11302, 7367**]

The fact that the sum of combinations equal the initial seed can be verified by pasting the "Combinations" expression into www.wolframalpha.com. This will provide the decimal (base-10) initial seed. To see it in hex form and verify it matches the key shown above, simply write it as:

BaseForm[*combination expression above*, 16]

### 7. The I.V. Salting Process

When used in a cryptographic context, in order to prevent statistical attacks, it is imperative any generated circular array never be reused when generating a bitstring. To this end, we employ an Initial Value ("IV"), which consists of 'R' random bits ideally derived from unbiased physical noise. The length of the IV can be as long or as short as is required by the implementation, but from a cryptographic perspective, we assert that the minimum Tumbler count should be 48 (due to the linkage with the Subset Sum problem, explained at the end of the paper), and thus the corresponding IV is 384 bits long; this specific number is designed to align with the least possible block key Tumbler count of 48 (=384 bits). Furthermore, if used in a strong cryptographic context with higher Tumbler count, we recommend a longer IV to suit the Tumbler count, and the ultimate strength of the system will be intrinsically linked to the quality of the generated IV.

As mentioned at the start of the paper, if used in a PRNG context, the use of another RNG to first construct the JP-Table and then to generate the IV is not redundant nor does it bias the random quality of the output, as the iterated nature of the system merely whitens any biasing effects of the initials RNGs.

The position of the IV salting process in the system will be described in the next section, here we simply describe it's operation. After a given Churn operation (noting that since the effect of the IV will change the input Tumbler Values, the result of the preceding Churn does not need to undergo "Repeated Tumbler Filtering"), the process of IV salting incurs using the XOR or ADC operations (hereafter we will keep with XOR for illustrative purposes) on the first Tumbler and the first 16-bit chunk of the IV, and the result put into the start of a new circular array, again the same size as the original JP-Table.

This process is repeated with the second Tumbler and the second 16 bit chunk of the IV. However this time, the result is compared against the first to ensure it did not yield the same result (a form of

Repeated Tumbler Filtering). Since this data will feed back into the machine, a non-unique result would have the same effect as a repeated Tumbler, yielding (a) XOR (a) = 0, reducing the effectiveness of the system by a factor of two Tumblers.

If the result is in fact the same as a previous result in this iteration, we disregard the current Tumbler, and move to the next, but remain on the current 16-bit chunk of the IV. This ensures the IV is fully utilized and not sacrificed for the slightly more deterministic outputs of the churning process. Once the IV has been fully exhausted, the remaining Tumblers are simply copied to the new circular array with no modification. If the IV is too large, there will simply be unused portions of it at the end of the salting process, and provides an indication that the IV is too long (it only needs to be as long as the Tumbler count in bits, and we assert that a 384 bit IV is more than sufficient to provide more cryptographic strength than is actually required for any practical application, although this is fundamentally a matter of implementation).

The process is shown pictorially as follows.

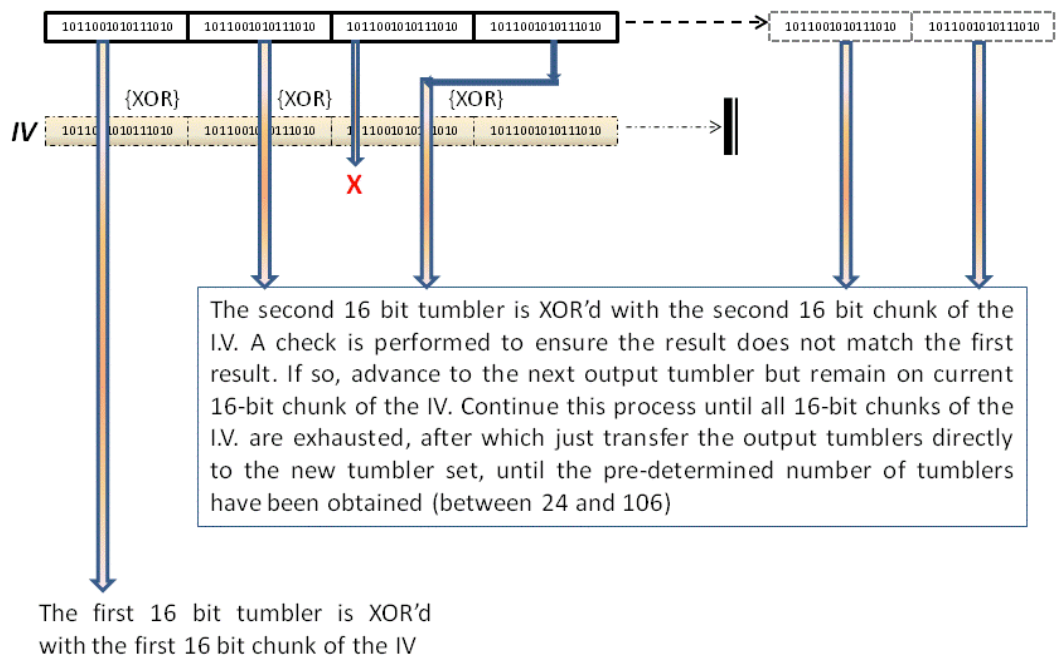

The first 16 bit tumbler is XOR'd with the first 16 bit chunk of the IV

The second 16 bit tumbler is XOR'd with the second 16 bit chunk of the I.V. A check is performed to ensure the result does not match the first result. If so, advance to the next output tumbler but remain on current 16-bit chunk of the IV. Continue this process until all 16-bit chunks of the I.V. are exhausted, after which just transfer the output tumblers directly to the new tumbler set, until the pre-determined number of tumblers have been obtained (between 24 and 106)

The result of the above process is another circular array, but this time not deterministic even in the forward sense due to the use of the IV. We have already asserted that it is not deterministic in the backward sense due to the destructive nature of the XOR or ADC operations, and to perform such a reverse-engineer of the system is tantamount to the NP-Complete Subset Sum problem.

## 8. Generating the bitstring

We now finally describe the process of generating the bitstring, that we claim is unique to the seed value and once produced, cannot be reverse-engineered to recover the initial seed value used to create it. The process is first described pictorially, using the churning symbol from Figures 2 and 3:

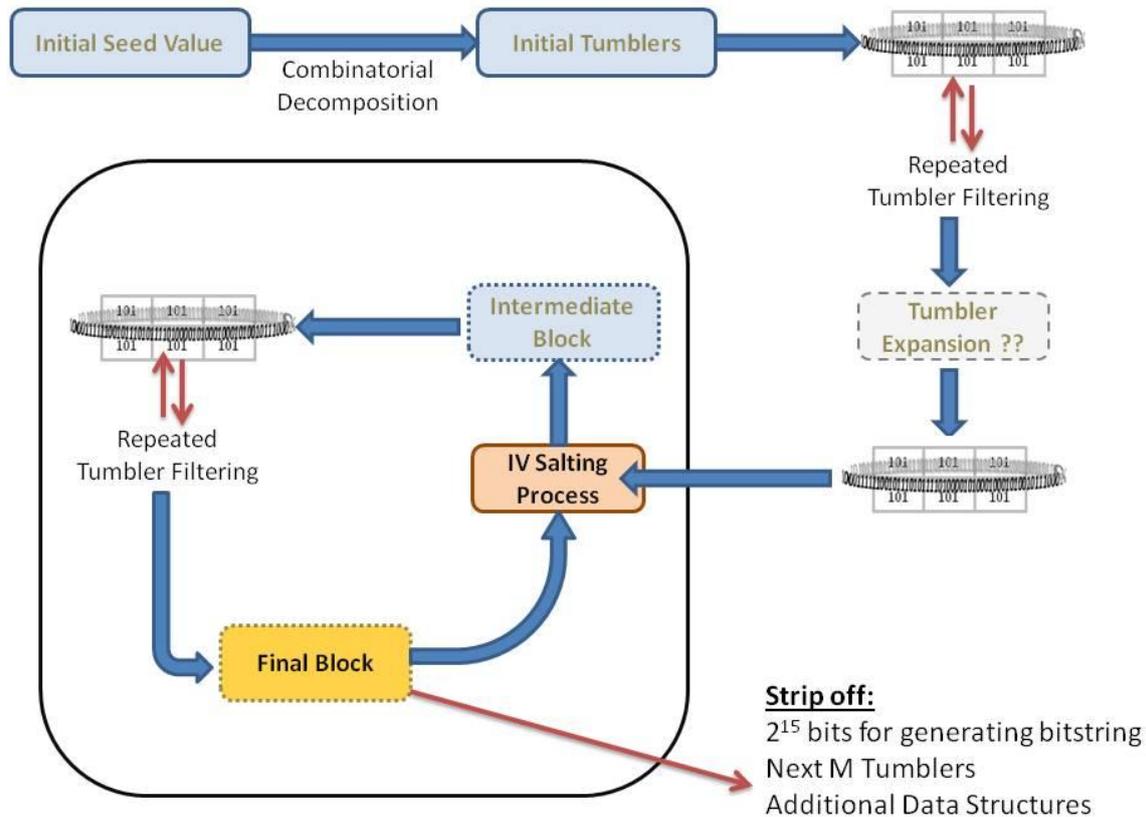

[Figure 4: The whole process]

The procedure is as follows, using the methodologies described in the preceding parts of this paper. We assume a circular array (a JP-Table) exists with randomly distributed 1's and 0's, and is $2^{16}$ = 65536 bits in size, and has a pre-determined bit position to be labelled as position "0", aka the "start point".

First, take the seed value (nominally but not necessarily $2^7$ to $2^9$ bits) as a single integer. The seed value could be a hexadecimal key or a password in the case of cryptographic applications, but is treated as a single integer nonetheless.

Using combinatorial decomposition, turn the seed value into a pre-determined series of numbers in the range 0 to [L – 1], where L is the table size (65536 bits in our application). Assume it has been determined that we need 'M' such numbers for our system to be effective.

Use these values to generate rotations of the original table, which could be anywhere from zero (the original table) to one complete rotation minus one bit position. Stack these rotations vertically , taking care to keep everything relative to bit position "0".

Perform a vertical XOR or ADC operation on each pair of tables to give CEIL(M/2) new tables. Repeat this until only one new table remains.

Taking this new table (circular array), count 16 bits from bit position zero (bits 0 to 15), and store this as a next new Tumbler Value. Then do the same for bits 16 to 31, bits 32 to 47, and so on in 16-bit chunks. For each new Tumbler recorded, ensure it does not match any previous tumblers. If it does, discard it and move to the next 16-bit chunk of the new circular array and continue. This is repeated until 'M' new Tumblers have been found.

At this point, optional Tumbler expansion can occur. This is desired when used in strong cryptographic contexts, in which one wishes to be certain that a brute force attack is still more efficient than the next-best-attack based on the Subset Sum problem. More information is given on this at the end of the paper; but it suffices to say that the Tumbler count at this stage can be anywhere between 10 and 106 Tumblers if the authors recommendations are taken. Tumbler Expansion simply occurs by continuing to read off new Tumbler Values beyond M to (M+d = M'), where M' is the new desired number of Tumblers.

At this point, we have generated what will be called the Constant Bit Mask. It is named as such as it represents a string of 65536 bits that depend solely on the initial seed value, and does not change provided the seed value does not change. At this point we enter the clearly marked "loop" section of Figure 4.

The bits of the Constant Bit Mask are then put through the IV salting process described in the previous section. This results in what we call the "Intermediate Block". This is then put through the churning process again, complete with checks for repeated Tumblers (Repeated Tumbler Filtering in Figure 4). This results in the "Final Block" which is a circular array of size 65536 that will be used to generate the required data.

Note there has been some debate amongst the authors as to whether this Final Block should be churned once more, with no checking for repeated Tumblers (as the output will be salted with the IV again before being put back through the state machine, which includes its own repeated result filtering). We conclude that whilst this increases the effective distance between the initial seed and the resulting bitstring, it takes time and does not provide demonstrable results, even from a cryptographic perspective. It is therefore left as a matter of implementation as to whether this additional step is taken.

From the resulting 65536 bits, the first half ($2^{15}$ bits) are copied off and appended to the running bitstring (or used to initialize it if this is the first pass through the loop structure). The reason for taking precisely half (and no more) of the available bits is twofold. First, it precludes cryptographic attacks on the systems by means of Gaussian Elimination, and second, it provides a great many more bits for other uses depending on the application. One necessary use is to generate the next M Tumblers, other uses could be to seed a cryptographic hash, or generate a second smaller bitstring for other uses separate from the primary bitmask.

After copying the first $2^{15}$ bits as the bitstring, we then start at bit position $2^{15}+1$, walking around the next M'+δ blocks of 16 bits to generate the next Tumbler set, where δ is the allowance for disregarding repeated Tumblers. In other words, the first $2^{15}$ bits should be used as the next 32,768 bits of the bitstring and nothing else – the following 32,768 bits should be used to generate the next set of M' Tumblers and any other useful data structures, such as extracting seeds to use in a cryptographically secure hash function.

### 9. Expanded Tumbler count and the Horowitz and Sahni adaptation

We have stated that when used in a cryptographic context, the Tumbler count should be expanded beyond that implied by the initial Tumbler set. We have also stated that at least a minimum of 10 Tumblers should be used. This arises from basic brute-force analysis of cryptosystems, and is a simply a recommendation when the scheme is implemented in a cryptographic context using 128 or 256 bit keys as seeds to the system.

We also recommend, for cryptographic security from the perspective that the system will be attacked from the Subset Sum perspective, The *expanded* Tumbler counts of [48-106] should be used. These seemingly arbitrary numbers come from the best known approach to the Subset Sum problem, namely the optimized Horowitz and Sahni adaptaion[4]. This ensures that a brute-force attempt to recover the initial seed value from a given subset of the bitmask remains more efficient than the next-best-attempt, which would be the optimized Horowitz and Sahni adaptation, described as follows:

Assume that the Tumbler count, T, is even. (The odd case is more complex, but has a solution time between the surrounding even cases, and thus offers no particular advantage.)

Produce a list of all possible newly formed circular arrays, several bits longer than T/2 Tumblers, which can be formed with T/2 Tumblers. There will be ($2^{16}$ nCr T/2) entries in this list. Along with the newly formed circular array, store its generating Tumblers. Create a copy of this list. In the original list, XOR every item with the presumed-known piece of newly formed circular array. Sort both lists, ascending. Walk up the lists at different rates, scanning for the first newly formed circular array which appears in both lists (but generally not at the same location). Inspect the T Tumblers, T/2 from each list, that were used to generate the equal newly formed circular arrays. Verify that no Tumbler is repeated. If so, continue the search.

It's now likely that the T Tumblers have been discovered. If the candidate plaintext fails statistical analysis, then just continue the search. If we neglect the sort, the operation count is ($2^{16}$ nCr T/2).

The above algorithm reveals all (T nCr T/2) ways of forming the correct T Tumblers from 2 sets of T/2 Tumblers. There are ways of statistically reducing the operation count based on the above fact; fundamentally, the order of the Tumblers is irrelevant.

At most, we can reduce the operation count by a factor of (T nCr T/2) based on the above fact. (This appears to be impossible, but it serves as a lower bound.) We ignore the cost of sorting, which might more than double the execution time.

The lower bound operation count is thus (($2^{16}$ nCr T/2) / (T nCr T/2)).

(The expected executed operation count might be half of that, which is probably compensated for by the sort, so we can take this formula as a pessimistic lower bound). It is this formula which gives rise to the fact that 10-to-43 initial Tumbler set Tumblers give rise to 24-to-106 Expanded Tumblers, respectively; essentially, we want to have enough block key Tumblers that to reverse them back from the newly formed circular array using this optimized cracking algorithm would be more complex than simply trying every (or rather, on average, half of) all possible initial Tumbler sets.

### 10. Limit cycles and improvements

Clearly a key characteristic of the system is the periodicity of the state machine. Or in other words, how long before we fall into a cycle of generating the same sections of the bitmask over and over again. This constitutes a failing both in a cryptographic context and a PRNG one, and described a limit of the system. The authors attempt to answer this by extrapolation of small data sets, and in doing so offer some possible improvements to the final system.

By taking small JP-Table sizes (up to 4096 bits) and small initial Tumbler counts (2-10) and running the system until it fell into a limit cycle, a series of histograms was produced to determine how many bits on average could be generated for various combinations of table size and Tumbler count, and how disperse the data set was. By extrapolating this data, we were able to show that with the suggested table size of $2^{16}$ bits and the minimum recommended 24 Tumblers, we could reliably generate $10^{27}$ bits of unique bitstring as a lower bound. Due to the large numbers involved this is all extrapolation as we lack the computation power to hard-test the limit cycle properties. However, we assumed that since the numbers were so large, it was of no great consequence.

Since this caused some discomfort in the community, we suggest also seeding the system using a linear congruential generator , aka a Marsaglian multiply-with-carry oscillator[5], which has a proven limit cycle of $10^{100}$. By using this to feed into the rotation of the JP-Table, there are no longer any (realistic) qualms about falling into a limit cycle, and adherence to the Subset Sum problem remains in terms of backtracking a chunk of the bitstring to the original seed value due to the XOR or ADC operations used on the stacks of circular arrays.

### 11. Conclusion

In conclusion, we have suggested a method of taking a seed integer value and converting to a long bitstring of useful length between $10^{27}$ and $10^{100}$ bits long (based on the limit cycle properties), that is also not reversible to recover the initial seed value.

We demonstrated this via taking that initial seed value, that could be a simple number, the output of another system, a cryptographic key or a RNG in its own right, and by way of combinatorial decomposition, followed by the suggested but optional use of a linear congruential generator (multiply with carry oscillator) with known periodicity, generated a series of integers in such a range as to provide unique and potentially all rotations of a pre-defined binary circular array. We then showed how these rotations can be combined via one-way functions (such as exclusive-OR and add-with-carry) to generate a bitstring unique to the seed value, and such that excluding a brute-force attempt, reversing the process is tantamount to solving the subset sum problem. We also show that additional strength is provided in a cryptographic context by means of salting the data during the iterative processes by use of a randomly generated "initial value".